# Deleterious satellite charging and possible mitigation schemes

Akash Yadav[1,2, a)] and S. K. Mishra[3, b)]

[1] Indian Institute of Science Education and Research Thiruvananthapuram, Thiruvananthapuram 695551, India
[2] National Institute of Technology Patna, Patna 800005, India
[3] Physical Research Laboratory (PRL), Ahmedabad 380009, India

[a)] akashjnvlko@gmail.com
[b)] nishfeb@gmail.com; sanjaym@prl.res.in (Corresponding Author)

---------------------------------------------

**Abstract:** Electrostatic charge dissipation is one of the major concerns for satellites operating in the earth's orbits. Under energetic plasma conditions, they may acquire very high negative potential (up to 10's of kV) due to the collection of energetic plasma constituents – resulting in temporary outages and permanent damages to onboard equipment. This study proposes and discusses a couple of physics-based schemes capable of mitigating/ minimizing the excessive charging effects over satellites under extreme plasma conditions in LEO/ GEO. An estimate of charge build-up on the space objects based on the charging dynamics as a function of ambient plasma parameters has been made. Our calculations illustrate that in the absence of a significant charge dissipation mechanism, a severe charging (~10's kV) in the dark/ shadowed at GEO and high latitude LEO regions. We propose that the installation of a suitable UV lamp and micro/nano-structuring of the surface fabric can induce efficient dissipation mechanism and effectively prevent the surface from the deleterious charging effects during satellite operation. We demonstrate that the UV illumination may maintain the satellite surface at quite a small positive potential (~ 2 V) while the surface nanofabrication sustains it at a sufficiently low negative potential (~ 10 V). Both concepts are shown to work efficiently in mitigating the potential threat of massive charging and safely performing the satellite operation.

-------------------------------

## 1. Introduction

Satellites in orbit are exposed to harsh space environments such as space debris, neutral and charged fine particles, and plasmas (Whipple, 1981). In particular, the interaction of satellite surfaces with the energetically charged plasma particles leads to the accumulation of charges on surfaces, i.e., termed satellite charging (Lai, 2019; Lai and Cahoy, 2016). The study of satellite charging has been a topic of interest due to the strategic and commercial importance of space and planetary satellites (DeForest, 1972; Garrett, 1981; Rosen, 2012). Excessive surface charging can lead to operational anomalies, physical damages to the satellites, or the complete loss of mission (Baker et al., 1994; Leach and Alexander, 1995). It is considered a severe concern for the ongoing and future orbiter missions (Pastena et al., 2020), including the satellites in earth's orbit campaigns.

The charging of the objects in space plasma in different scenarios, for Maxwellian and Non-Maxwellian plasmas, has been widely discussed in the literature (Lhotka et al., 2020; Mishra and Misra, 2015; Shukla and Mamun, 2002; Whipple, 1981). Due to higher mobility, the electron current dominates the charging process (Sodha, 2014) and results in negative charging of the satellite during eclipses (Thomsen et al., 2013). Satellites in the geosynchronous earth orbit (GEO) experience large charging effects due to the presence of highly energetic (~10 keV) plasma (Matéo-Vélez et al., 2018). The analysis of charging events of geosynchronous satellite ATS reveals that satellite surfaces can acquire a very high negative potential of ~10 kV (DeForest, 1972). Although the low latitude region of LEO is relatively safe, satellites orbiting in highly inclined LEO encounter high negative charging while passing through the auroral region (Colson and Minow, 2011; Gussenhoven et al., 1985; Meng et al., 2017). Analysis of the charging events of DMSP spacecraft shows that the satellite can attain negative potential up to 2 kV during the auroral activity (Anderson, 2012). The electric fields from charged surfaces may disturb the local plasma environment (Lai, 2019), interfere with the onboard sensitive scientific instruments and cause errors in the scientific measurement (Lubomudrov et al., 2019).

In case the object is illuminated by sunlight, the photoelectron emission contributes significantly to the charging and effectively reduces the negative charge; the dominant photoemission flux (over electron collection) might lead to a positive potential on illuminated satellite surfaces (Misra et al., 2015). Satellite consists of several components that may not be electrically grounded. In the case of partial illumination, the shaded surfaces develop high negative potential, whereas the illuminated surfaces may attain positive potential – this might lead to differential charging (Lai and Tautz, 2006). When the potential difference between these surfaces exceeds breakdown voltage, electrostatic discharge (ESD) occurs in the form of arcs. An ESD can damage essential electronic components, circuits, sensors, and power systems (Gupta et al., 2015; Rosen et al., 1978). ESDs can also cause physical surface damage and degradation of satellite material (Lai and Cahoy, 2016), E.g., on January 20, 1994, the geosynchronous communication satellite Anik E-1 experienced high electrostatic charging where the resulting ESD damaged the gyroscopic guidance system of the satellite. Due to the failure of the guidance system, the satellite started an uncontrolled wobbling motion and caused hours of disruption in television, telephone, and computer data transmissions (Koons et al., 1998; Lam et al., 2012). The geosynchronous satellites Anik E-2, Intelsat-K have also shown a similar anomaly. These are some representative cases of satellite anomalies developed due to excess satellite charging – hundreds of such satellite anomalies and failures have been documented (Koons et al., 1998; Leach and Alexander, 1995).

To address this serious issue, we have adopted a physics-based approach to model satellite charging and demonstrate effective mitigation methods to tackle the deleterious effects of excessive satellite charging. Satellite charging is a complex process that not only depends on the ambient environment but also on the satellite surface material and the geometry of the satellite (Fennell and Anderson, 2008). For the sake of simplicity in the analysis and illustration of the concept, we have treated the satellite as a metallic sphere, and the ambient plasma is assumed to follow Maxwell-Boltzmann velocity distribution. With these simplifications, the theory of surface charging in plasma is extended to satellite charging (Whipple, 1981). The rate of accretion of the plasma particles on the satellite surfaces is calculated using the OML (orbital motion limited) approach (Sodha, 2014), whereas Fowler's approach to photoelectron emission is used for photoelectron flux calculation (Sodha, 2014). With this model, the satellite charging due to ambient plasma environment and solar irradiation corresponding to ionospheric and magnetospheric regions associated with LEO and GEO is calculated. The calculation results are in agreement by order of magnitude with the charging observed in the previous missions (Anderson, 2012; Francis, 1982; Gussenhoven et al., 1985; Matéo-Vélez et al., 2018). They could be detrimental to the routine operation of active satellites.

To overcome this concern, traditional approaches rely on strict design guidelines, e.g., shielding, material selection, and grounding components, to avoid differential charging (Garrett et al., 2012). Although grounding reduces the risk of differential charging, it does not mitigate the excess charge from the satellite structure. Recent studies have suggested electron emission-based approaches to mitigate the surface charge (Iwata et al., 2012; Kureshi et al., 2020; Lai, 2012; Mishra and Bhardwaj, 2020; Mishra and Sana, 2022). These mitigation techniques are quite effective and well understood, but have not been yet explored in-depth in the context of satellite charging. In this paper, we have demonstrated two possible mitigation schemes, viz., by (i) installation of an UV lamp operative during the eclipse, i.e., when the object is in dark/ shadow, and (ii) micro/nano-fabrication of the surfaces. The notion of the first scheme comes from the charging effects in illuminated conditions – here, UV lamp illumination might trigger the photoemission from the satellite surface, which can be a potential discharging mechanism to mitigate the excessive negative charge from the satellite during the eclipse. The other scheme is conceptualized from the fact that the micro/nano-structures on the fabric of the satellite might act as field emission centres (Mishra and Bhardwaj, 2020). It can induce sufficient field emission (F.E.) current to neutralize the excessive negative charge development over satellites traversing the eclipse zone (Mishra and Sana, 2022). We have conceptualized both mitigation methods and performed simulations to verify the significance of these mechanisms in neutralizing the detrimental satellite charging effects.

**2. Surface charging**
To illustrate the conceptual basis, we first establish the analysis deriving the charge development over the surface of orbiting satellites. The surface charge at any location in orbit depends on solar radiation and the ambient plasma environment. Orbiting around the earth, the satellite encounters different plasma compositions and solar radiation. For instance, in the illuminated regime, the satellite faces typical solar wind (sw) plasma environment, while in eclipse (dark), the satellite generally interacts with diffused sw plasma. The temporal evolution of charge can be expressed in the form of the current balance equation (Lhotka et al., 2020; Mishra and Misra, 2015).

$$\frac{dQ}{dt} = I_e + I_i + I_{ph}$$

(1)

where $Q$ is the satellite charge at any instant, $I_e$ is the current due to electrons, $I_i$ is current due to ions, $I_{ph}$ represents the photoemission current.

For simplicity in the analysis, we limit the study to spherical objects (satellite) and Maxwell-Boltzmann velocity distribution of the sw plasma population. The constituent electron accretion flux (Sodha, 2014), in a plasma environment, can be expressed as,

$$I_e^+ = -q_e n_e \pi a^2 \sqrt{\frac{8k_B T_e}{\pi m_e}} (1 + \alpha_e Z)$$

$$I_e^- = -q_e n_e \pi a^2 \sqrt{\frac{8k_B T_e}{\pi m_e}} (\alpha_e Z)$$

(2)

and ion accretion flux, in a plasma environment, can be expressed as,

$$I_i^+ = q_e n_i \pi a^2 \sqrt{\frac{8k_B T_i}{\pi m_i}} \exp(-\alpha_i Z)$$

$$I_i^- = q_e n_i \pi a^2 \sqrt{\frac{8k_B T_i}{\pi m_i}} \exp(1 - \alpha_i Z)$$

(3)

where, $q_e$ is the charge of the electron, $Z = Q/q_e$ is the charge number, $n_e$ and $n_i$ represent the electron and ion density in the ambient plasma, $T_e$ and $T_i$ represent electron and ion temperature in the plasma, respectively, $k_B$ is Boltzmann constant, $m_e$ and $m_i$ represent the mass of electron and ions, respectively, and $a$ represents the radius of satellite. The superscript $(+, -)$ represents a positive or negatively charged surface. This notation is followed throughout the paper. Here, the constants $\alpha_e$ and $\alpha_i$ are given by,

$$\alpha_e = \frac{Z q_e^2}{4\pi \epsilon_0 a k_B T_e}$$

and,

$$\alpha_i = \frac{Z q_e^2}{4\pi \epsilon_0 a k_B T_i}$$

When the satellite passes through the illuminated region, the current resulting from photoelectric emission plays a vital role in the charging process. The photoemission flux ($n_{ph}$) from the spherical surfaces (Sodha, 2014), based on Fowler's approach (Fowler, 1955) to photoelectron emission, for the negatively charged surface can be expressed as

$$n_{ph}^- = \int_{v_{min}}^{v_{max}} \chi(v) \Lambda(v) dv$$

In the case of a positively charged surface, the photoemission flux can be expressed as

$$n_{ph}^+ = \int_{v_{min}}^{v_{max}} \chi(v) \left[\frac{\psi(\xi, (Z+1)\alpha_s)}{\Phi(\xi)}\right] dv$$

(4)

where, $\nu$ is the frequency of the incident photon, $\chi(\nu)$ is the photoelectric efficiency – it is the ratio of the number of photoelectrons emitted per incident photon, $\Lambda$ is the incident photon flux, and ($\nu_{min}, \nu_{max}$) is the frequency range of the solar spectrum, and $\psi$ is given by,

$$\psi(\xi, (Z+1)\alpha_s) = (Z+1)\alpha_s \ln[1 + \exp(\xi - (Z+1)\alpha_s] + \Phi(\xi - (Z+1)\alpha_s)$$

The constant $\alpha_s = Zq_e^2/4\pi\epsilon_0 ak_B T_s$, value of $\Phi(\xi) = \int_0^{\exp(\xi)} \kappa^{-1}\ln(1+\kappa)\,d\kappa$ is obtained by numerical integration, the upper limit depends on $\xi$, which is defined as $\xi = (h\nu - \phi)/k_B T_s$, $\phi$ is the work function of the surface material, $h$ represents the Planck's constant, and $T_s$ is the satellite surface temperature.

For the photoemission, the energy of the incident photon should be greater than the work function($\phi$) of the surface material; therefore, the minimum frequency for photoemission can be expressed as, $\nu_{min} = \phi/\hbar$. Following Draine's formulation (Draine, 1978), the spectral dependence of the photoelectric efficiency can be expressed as, $\chi(\nu) = \chi_0(1 - \nu_{min}/\nu)$. In this expression, $\chi_0$ is the photoelectric yield, a measure of the fraction of photon flux contribution to photoelectric emission flux.

Eq.1 for satellite charging for the negatively (Eq.5) and positively (Eq.6) charged surface can be rewritten as

$$\frac{dz_n}{dt} = \alpha_n \left[ \left(1 - \frac{T_e}{T_i} z_n\right) + \sqrt{\frac{\pi}{8}} \frac{n_{ph}^-}{n_i v_i} - \frac{n_e v_e}{n_i v_i} \exp(z_n) \right] \tag{5}$$

$$\frac{dz_n}{dt} = \alpha_n \left[ \exp\left(-\frac{T_e}{T_i} z_n\right) + \sqrt{\frac{\pi}{8}} \frac{n_{ph}^+}{n_i v_i} - \frac{n_e v_e}{n_i v_i}(1 + z_n) \right] \tag{6}$$

where, $z_n = Zq_e^2/4\pi\epsilon_0 ak_B T_e$ is the normalized charge number (w.r.t electron temperature), the constant $\alpha_n = n_i a q_e^2 v_i/\sqrt{2\pi}\,\epsilon_0 k_B T_e$, $v_e = \sqrt{k_B T_e/m_e}$ and $v_i = \sqrt{k_B T_i/m_i}$ are the thermal velocity of the electrons and ions respectively. The normalized charging equations Eq. (5) and Eq. (6) apply to both the dark and sunlit regions. In the case of the dark region, the photoelectron emission flux becomes zero, so the second term of the equation vanishes.

## 3. Surface charge calculations
### 3.1 Computational approach and data

The temporal evolution and steady-state characteristic of the surface charge or potential on the satellite at any given location in its orbit are determined by solving the current balance equation (Eq.1) with the appropriate sw plasma and solar irradiation parameters and charging currents. We have used in-built differential equation stiff and non-stiff ode solver MATLAB packages – ode15s, ode23s, and ode23t to solve the charging equation. We use uncharged surfaces, i.e., z = 0, at $t = 0$, as the initial condition. As discussed earlier, satellites encounter a variety of plasma environments depending on their altitude and latitude in orbit. The ambient plasma characteristics can be described in terms of electron number density ($n_e$), electron temperature ($T_e$), ion number density ($n_i$) and ion temperature ($T_i$) within the given plasma population. The computations are performed for most of the possible scenarios the satellites encounter in their orbits around the earth.

In the GEO region, the plasma density is $\sim 1 \text{ cm}^{-3}$ and electron energy is of the order of tens of keV (Francis, 1982). The geostationary plasma environment is very dynamic and energetic, especially during the geomagnetic substorms. The injection of highly energetic particles from magnetotail can raise the electron temperature to $\sim 13\text{KeV}$ and ion energy to $\sim 20\text{KeV}$ (Rosen, 1976). The plasma density during this event is $\sim (1.2 - 1.6) \text{ cm}^{-3}$. NASA's worst-case plasma parameters (Anderson and Smith, 1994), based on the observations from earlier space missions, are as follows: electron density $\sim 1.12 \text{ cm}^{-3}$, ion density $\sim 0.236 \text{ cm}^{-3}$, electron temperature $\sim 12\text{KeV}$ and ion temperature $\sim 29.5$ KeV. In the Plasma sheet, the plasma parameters are: density $\sim 1 \text{ cm}^{-3}$, electron energy 1 KeV and ion energy $\sim 6\text{KeV}$ (Grard et al., 1983). Under extreme sw conditions, the magnetopause is pushed inside GEO, exposing the satellites to the magnetosheath plasma (Shue et al., 1998). This event is known as geosynchronous magnetopause crossing (GMC), and it is quite rare – almost 100 such cases have been reported over 20 years (Samsonov et al., 2021). In magnetosheath, the plasma density refers to $\sim 30 \text{ cm}^{-3}$ while electron/ion energy is $\sim 80$ eV (ECSS, 2020). The low latitude LEO plasma is dense ($10^5 \text{ cm}^{-3}$) and cold ($\sim 0.1$ eV) compared to the GEO region, but at high latitudes, orbits receive high energy electron precipitation resulting in auroral activity. During this event, the electron density and temperature correspond to $0.1 \text{ cm}^{-3}$ and 400 eV, respectively (Kletzing et al., 2003).

The solar spectrum data from ASTM E490-00, Extra-terrestrial Spectrum Reference (ASTM, 2000) is used to calculate the photoemission current from the satellite surface under sunlit conditions. The solar spectrum is shown in Fig.1. The photoemission depends on the electronic properties of the material used in manufacturing satellite surface, i.e., work function, operating temperature, and photoelectric yield. Commonly aluminium and its alloys are used for satellite design. Therefore, we have simulated satellite charging corresponding to aluminum ($\phi = 4.06$ eV) for the illuminated conditions. We have performed calculations for the satellite temperature $T_s = 255K$ (Meseguer et al., 2012) and photoelectric yield $\chi_0 = 10^{-4}$ (Feuerbacher and Fitton, 2003). As mentioned earlier, the satellite is assumed to be a sphere of a 1-meter radius for numerical calculations. In reference to the charging equation (Eq.1), note that the transient evolution of the surface potential ($V = ze/a$) depends on the size of the spherical object, but the steady state values are independent of its size (Sodha et al., 2010) – it suggests that the steady state surface charge varies linearly with the radius of the spherical object.

These discussed plasma parameters and solar spectra for various physical conditions are used to perform calculations for the satellites orbiting in LEO and GEO orbits – the charging results are discussed next.

**3.2 Surface charge estimates**

First, we discuss the charging in the dark region. Such a situation may correspond to the satellite in the night segment of orbit or the shaded region of the partially sunlit satellite. In this case, the satellite surface interacts only with the ambient plasma and acquires charge. Being lighter than the ions, electrons show higher mobility and larger flux towards the surface compared to the ions. Hence, the satellite surface generally acquires a negative charge in the dark regions. As the surface charging equation (Eq.1) suggests, the electron current and ion current become equal, i.e., $I_e = I_i$, in the steady state.

The numerical simulation results for charging in the dark region under different plasma conditions are shown in Fig.2. We note, under distinct conditions, the satellite gets charged up to: during geomagnetic substorm ∼(−30 kV), in the auroral region ∼(−1.2 kV), in plasma sheet ∼(−2.5 kV), in magnetosheath ∼(−200 V), solar winds ∼(−35 V), plasmasphere ∼(−2.5 V) and low latitudes LEO ∼(−1.6 V). The surface may acquire ∼ (−4 kV) potential for a typical GEO region with average plasma parameters. Using NASA's standard worst-case plasma parameters, we show that the satellite surface can charge up to ∼(−40 kV). Such a high potential in the ∼kV range might be detrimental to the functioning of electronic circuits/ components. Fig.3 represents the surface potential of the satellite over sunlit locations. In this case, surface charging occurs due to the interplay of the flux from ambient plasma and photoelectrons. The ambient plasma accumulates a negative charge on the satellite surface, while photoemission releases the electrons – the competing effect gives rise to the charge on the satellite surface. For the moderate plasma conditions and solar radiation spectrum, e.g., LEO plasma parameters, the photoelectric emission flux usually dominates over the accretion flux from ambient plasma, and the surface acquires a positive potential of a few volts. From the simulation results, we may conclude that in the presence of solar radiation or illuminated conditions, the surface potential is quite small (<10 V) and should not cause significant damage to the satellite.

**4. Mitigation of Excessive Surface Charge**

The numerical results in the last section indicate that the satellite can acquire very high negative potential, particularly in the dark (eclipse) region, which might yield severe damage to its functioning. These damages may cause the temporary or permanent failure of the components of the satellite and, in the worst case, can even lead to mission termination (Koons et al., 1998; Leach and Alexander, 1995). Hence, it is necessary to mitigate such excessive charge deposition on satellites. Following, we discuss the conceptual basis of two possible mechanisms that could efficiently source excessive charge mitigation.

**4.1 Implantation of an UV lamp**

We observe severe charging effects in the dark region, while nominal/ moderate charging is noticed over the sunlit locations. In fact, in the case of sunlit surfaces, the photoelectric emission naturally mitigates the excess negative charging created by ambient plasma sources. It suggests a novel notion that the excessive negative charge could be mitigated by shining the surface with an UV source (lamp) of suitable frequency and intensity.

A schematic illustrating the conceptual basis is depicted in Fig.4. To ensure the photoelectric emission from the surface, the frequency of the UV lamps should be greater than the threshold frequency of the surface material. To illustrate this concept, we use the charging analysis for the dark region by including an additional monochromatic UV source capable of generating the photoelectrons from the satellite surface. The flux connected with the photoemission due to the monochromatic UV lamp of frequency $\nu_{uv}$ corresponding to negatively and positively charged surfaces, respectively, can be expressed as

$$n_{ph}^- = \chi(\nu_{uv})\Lambda(\nu_{uv})$$

$$n_{ph}^+ = \chi \left[ \frac{\psi(\xi, (Z+1)\alpha_s)}{\Phi(\xi)} \right]$$

(7)

After implementing this mitigation scheme, this additional charging effect is manifested with the charging equation (Eq.1) for the dark region to evaluate the surface charge/ potential. For illustration, the case of satellite charging in the GEO region is calculated and illustrated in Fig.5. The calculations are made for $\nu_{uv} = 2 \times 10^{15}$ Hz, photo-efficiency $\chi_0 = 10^{-4}$ and the standard set of parameters discussed in Section 3.1 for different values of UV lamp intensity. It should be noted that the schematic shown in Fig. 4 is for illustration purpose. In real scenario the uniform illumination is desired to avoid differential charging. Hence, multiple UV light sources may be considered depending on the satellite geometry. The UV lamp intensity in our analysis refers to the intensity of UV radiation received at the surface of the satellite. To illustrate its contrasting behaviour, the case with zero intensity of UV lamp (absence of the illumination) has been included (red line in the bottom panel) – in this case, the surface acquires a negative potential of ~8 kV.

As the intensity of the UV lamp (viz., photon flux) is increased to $0.1$ Wm$^{-2}$, $0.5$ Wm$^{-2}$ and $0.7$ Wm$^{-2}$ the negative surface potential reduces to 6 kV, 2 kV, and 0.5 kV, respectively. If the intensity is further increased to $1$ Wm$^{-2}$, the surface acquires small positive potential of ~ 2V – it puts the satellite surface in a safe regime and avoids any detrimental effect from electrostatic charging. A further increase in the UV lamp intensity, for instance, $10$ Wm$^{-2}$ slightly raises the surface potential to ~ 4V. We note the increase in power does not alter positive potential much, but the steady state is reached at a shorter time scale. This could be understood in terms of increasing photons with increasing lamp intensity which effectively produces photoelectrons at a faster rate, and the plasma accretion current could be compensated at a shorter time scale. Other way around, a positive charge build-up on the surface increases the barrier height for electron emission. Hence, a small variation is observed in the steady state potential values. The effect of UV source intensity on the steady state surface potential for different $\chi_0$ values are shown in Fig.6. The variation in surface potential with UV lamp intensity follows a similar trend as discussed earlier in Fig.5. The increase in surface potential with $\chi_0$ may be attributed to the increasing rate of photoemission flux.

From our calculations, it may be concluded that UV light illumination of the surface of sufficient intensity ($1$ Wm$^{-2}$) neutralizes the disastrous electrostatic charging and acquires a small positive potential in ms (millisecond) time scale – the effect is attributed to the induced photoemission flux from the illuminated surface. During the operation, the intensity and spectrum of the UV source, if applicable, could be tuned to the desired extent depending on the ambient plasma characteristics and subsequent charging effects. For instance, higher-intensity illumination can be more effective in the case of highly energetic ambient plasma conditions (e.g., during substorms).

**4.2 Micro/nano-structuring of the satellite fabric (surface)**
Another scheme of charge mitigation is based on the phenomenon of electric field emission (F.E.) of the electrons. We propose that micro (nano) structuring of the surface of the satellite surface can efficiently support the charge dissipation and mitigate the massive charge deposition – a schematic of the concept is presented in Fig.7. The physics basis of the concept is as follows: a large negative potential (Fig.2) might create intense electric field over the tips of nano/micro-structures. It can trigger the electric field emission of low energy electrons via the quantum tunneling effect to facilitate the neutralization current and compensate for the accreting sw plasma flux. The FE electron flux from the negatively charged surfaces can be expressed as (Mishra et al., 2012),

$$n_{fe} = \frac{A_0 T^2}{q_e} \int_{e_r=w_a-v_s}^{w_a} \int_{e_t=0}^{\infty} T_p(e_r) f_d(e_r + e_t - e_f) de_r de_t \tag{8}$$

where, $A_0 = 4\pi q_e m k_B^2 / h^3$, $w_a = W_a/k_B T$, $e_r$, $e_t$ and $e_f$ are the radial and tangential electron energy (normalized with $k_B T$), $T$ is the temperature of the metal lattice, $v_s = V_s/k_B T$ is the potential energy of an emitting electron near the charged surface and $f_d = [1 + \exp(e_r + e_t - e_f)]^{-1}$ represents the Fermi-Dirac (FD) distribution (Seitz, 1940). $T_p(e_r)$ refers to the tunneling probability and can be expressed as (Ghatak and Lokanathan, 2004),

$$T_p(e_r) = \exp\left[-2\beta \int_1^{\rho_0} \left(\left(\frac{v_s}{\rho}\right) - (e_r - w_a + v_s)\right)^{1/2} d\rho\right] \tag{9}$$

where $\rho = r/a_0$, $r$ is the radial distance, $\beta = a_0 \sqrt{2 m_e k_B T}/\hbar$, $m_e$ is the mass of the electron, $\rho_0 = v_s/(e_r - w_a + v_s)$ and $\phi = W_a - E_f$ is the surface work function.

To illustrate the concept, we consider the field emission points are spherical tips (grains) of radius $a_0$ and the satellite surface is nanofabricated as shown in Fig.7. Considering the emission tips are hemispherical, then the total number of emission tips on the satellite surface is $4a^2/a_0^2$, and field emission current from each tip is $2\pi a_0^2 q_e n_{fe}$. The net field emission current from the satellite surface can be written as, $I_{fe} = 8\pi a^2 q_e n_{fe}$. Considering the contribution of F.E. current along with sw (electrons and ions) collection over the satellite surface for negatively charged surface, the normalized charging equation can be expressed as

$$\frac{dz_n}{dt} = \alpha_n \left[1 - \frac{T_e}{T_i} z_n + \sqrt{8\pi} \frac{n_{fe}}{n_i v_i} - \frac{n_e v_e}{n_i v_i} \exp(z_n)\right]. \tag{10}$$

Accounting for this effect, i.e., the inclusion of nano-structuring, the temporal evolution of the surface charge/ potential corresponding to the geomagnetic substorm is illustrated in Fig.8 for $W_a = 15.76 \, eV$, $E_f = 11.7 \, eV$ (Sodha, 2014), standard data discussed in Section 3.1 and different values of tips radius. It is noticed that the inclusion of the field emission sufficiently reduces the negative charging and brings the surface potential to a tolerable regime where the setup operates efficiently. The temporal evolution can be understood as follows: The charging begins with dominant sw plasma collection with a negligible FE contribution. As the surface acquires negative charge, a Coulomb potential builds up and leads to the FE current. This effect is pronounced for the spherical tips. The increase in surface potential increases the FE contribution. As soon as the tips acquire sufficient negative potential to generate sufficient neutralizing FE current, the surface reaches a steady state potential, as depicted in Fig.8.

The dependence of steady state surface potential on the tip size corresponding to substorm and high latitude plasma conditions is illustrated in Fig.9. It suggests smaller (in magnitude) surface potential in case of reducing tip size (radius). For the micron ($\mu m$) size tips, a significant negative potential, i.e., $\sim(1-10)$ kV on the surface is observed. In contrast, as the tip radius is reduced to the nm range, the potential reduces to a sufficiently small value $\sim(10-100)$ V within the time scale of microseconds. It can be understood in terms of the increasing intensity of the induced electric field with decreasing tip radius, which consequently increases FE flux and rapidly dissipates the excess accumulation of negative charge. The fabric of the satellite can be customized for specific cases, i.e., for GEO satellites tip radius should be in the nanometre ($1\ nm - 10\ nm$) range due to extreme charging events. These calculations demonstrate that the nanofabricated satellite surfaces can naturally mitigate/ minimize the excessive negative charging to an operational limit, even during the geomagnetic substorms where it shows massive $\sim$10 kV electric potential on the orbiting satellite.

**5. Summary and Discussion**
The excessive surface charging of satellites in the earth's orbit due to interaction with ambient plasma has been a serious concern and is considered responsible for several anomalies and failures in satellite functioning. With our calculations, we notice that the satellite operating in the eclipse zone is seen to gain high negative surface charge in GEO (few kV to ~40 kV) and high latitudes LEO (~1.2 kV) regions. The field originating due to this massive charging can interfere with the sensitive scientific measurement instruments and might overshoot the threshold, damaging electronic devices and setup. Moreover, severe differential charging can result in arcing, which can damage electronic devices, solar panels, and other onboard instruments.

To tackle this problem, in this study, we suggest and conceptualize a couple of feasible solutions to mitigate/ minimize the deleterious charging effects of space plasma over orbiting satellites. One of the solutions is based on photoemission, and we propose the installation of an UV lamp over the operating object. In the eclipse zone, when the satellites face the highest risk of severe negative charging, the UV lamp tuning might cause the photoemission of electrons and consequently suppresses the magnitude of the negative charge. Through simulation, we demonstrated that under the influence of an UV lamp, the surface acquires a finite but small positive charge keeping the satellite out of any potential threat. For instance, an UV lamp with intensity ~1 Wm$^{-2}$ maintains the operating surface at 2V under substorm conditions. The other physical mechanism we propose is a natural consequence of surface morphology. We suggest that micro (nano) structuring of the surface of the satellite fabric (surface) may act as the natural source of charge dissipation through field emission where these micro (nano) tips act like FE centres and maintain the surface at a significantly lower electric potential. This process has the advantage that being a natural process, it does not require any additional power. Our calculations suggest that the surface fabricated with 10 nm spherical tips can maintain the satellite surface at ~10 V negative potential in contrast to natural charging of ~10 kV (during geomagnetic substorms). We conclude that both the schemes have potential utility to efficiently dissipate the excess surface charge and maintain the charge of the operating surface at a safe level.

However, the surface charging estimation and efficiency of mitigation methods is done using a simplistic model, but it captures the underlying physics of electric charge/potential development on space objects and subsequent mitigation schemes. This study could be a basis for a more rigorous analysis of satellite charging and mitigation methods including more realistic conditions like satellite geometry/ local plasma environment and effect of other physical mechanisms (like secondary electron emission, plasma wake, etc.) which can be a direction of further investigation.

The proposed mitigation techniques can be tested in controlled realistic space plasma conditions to validate the efficiency and investigate the possible artifacts. Some possible challenges may involve surface damage due to erosion and sputtering due to bombardment of energetic particles, leading to reduced emission efficiency over time. In the case of UV lamp implementation, ensuring uniform illumination across satellite surface may impose extra certain design constraints. These issues need to be rigorously analyzed and optimized design with more particulars should be considered for real-world applications.

**Acknowledgment**

This work is supported by the Department of Space, Government of India.-----------------------------------------

----------------------------------------------------------

**Fig.1**

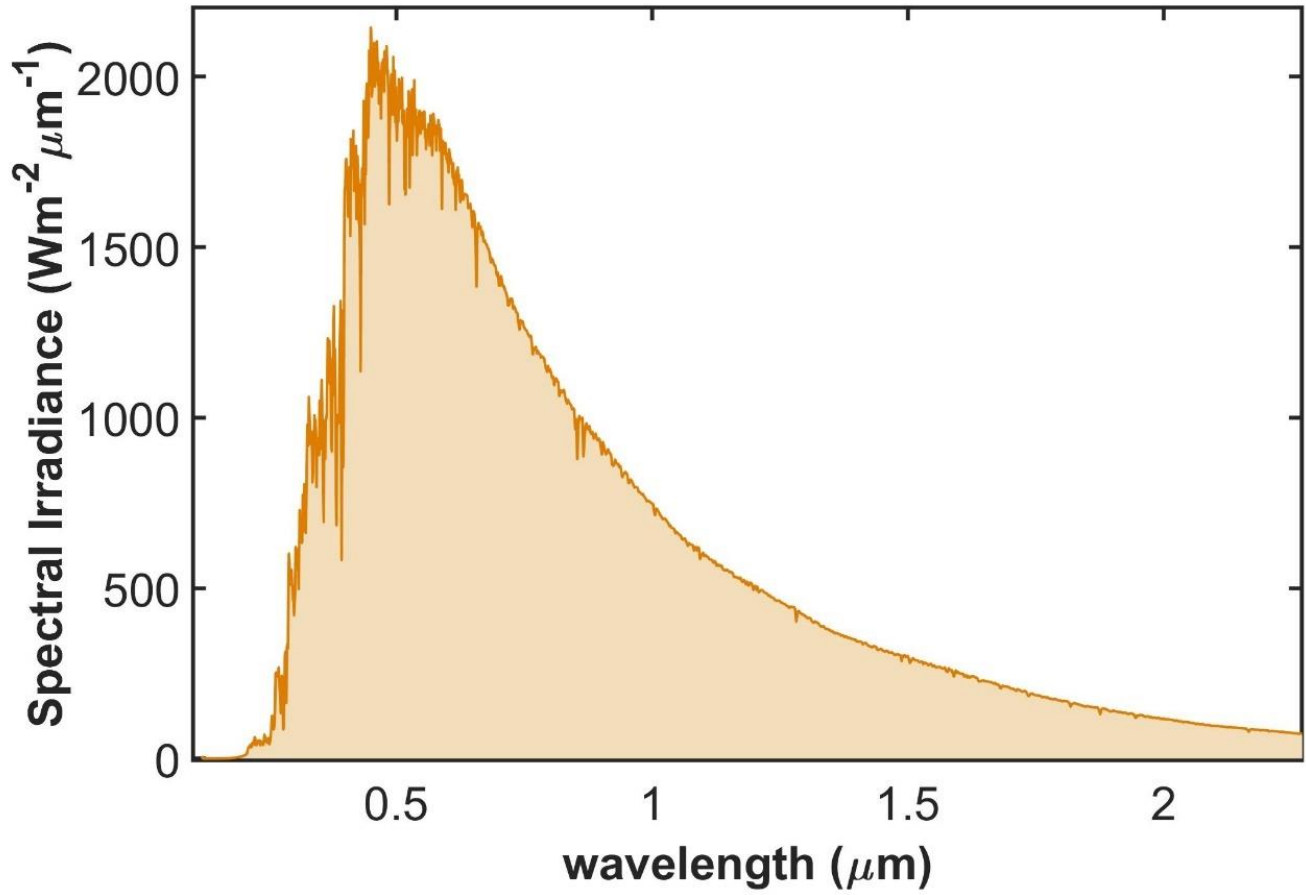

**Fig.1:** Extraterrestrial Solar Spectrum (*ASTM E490-00 Extra-terrestrial Solar Spectrum*).

**Fig.2**

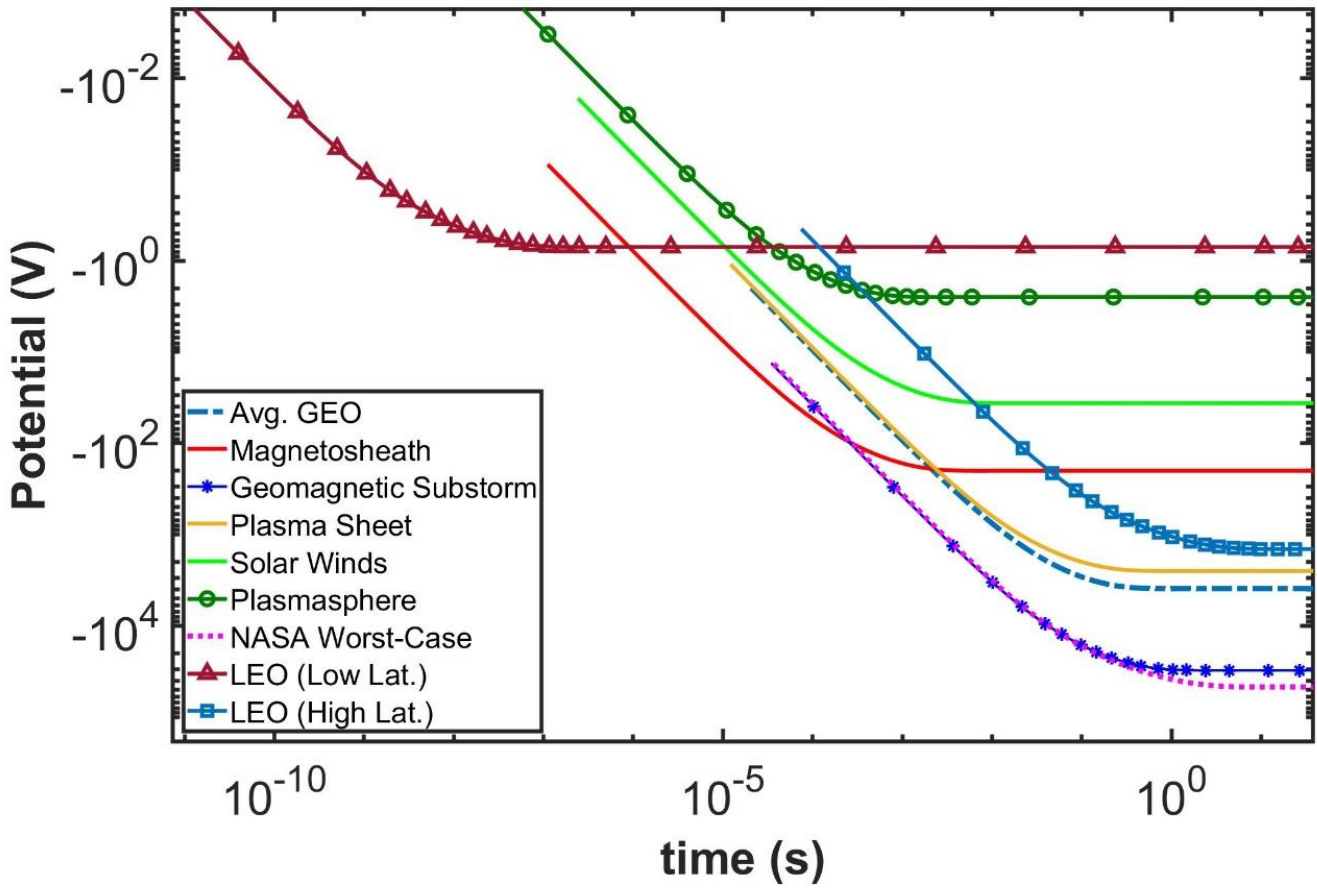

**Fig.2:** Temporal evolution of the surface potential of the satellite operating in the dark/ shaded region of LEO/ GEO under different plasma conditions (marked with colored labels).

**Fig.3**

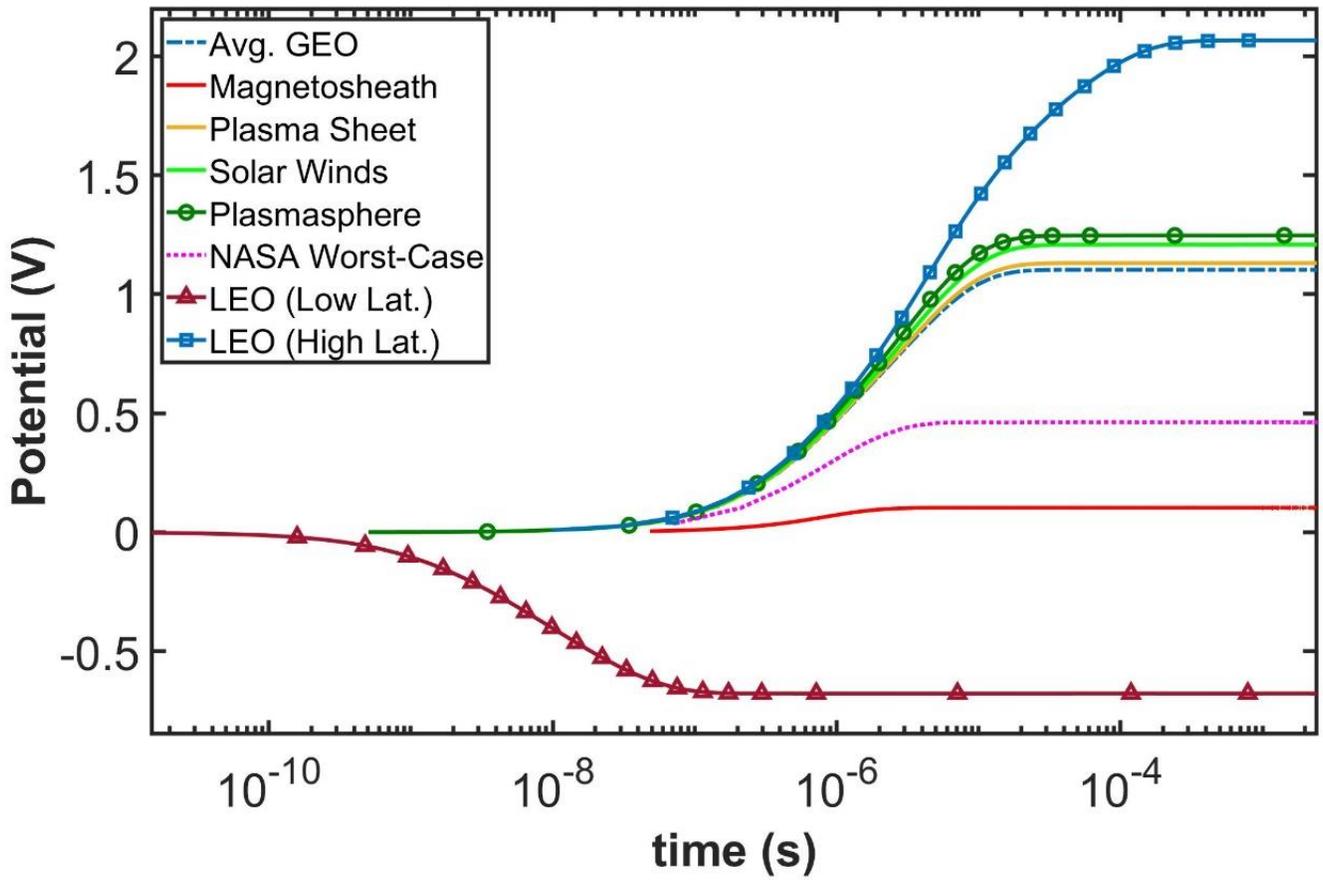

**Fig.3:** Temporal evolution of the surface potential of the satellite operating in the sunlit region of LEO/ GEO under different plasma conditions (marked with colored labels).

**Fig.4**

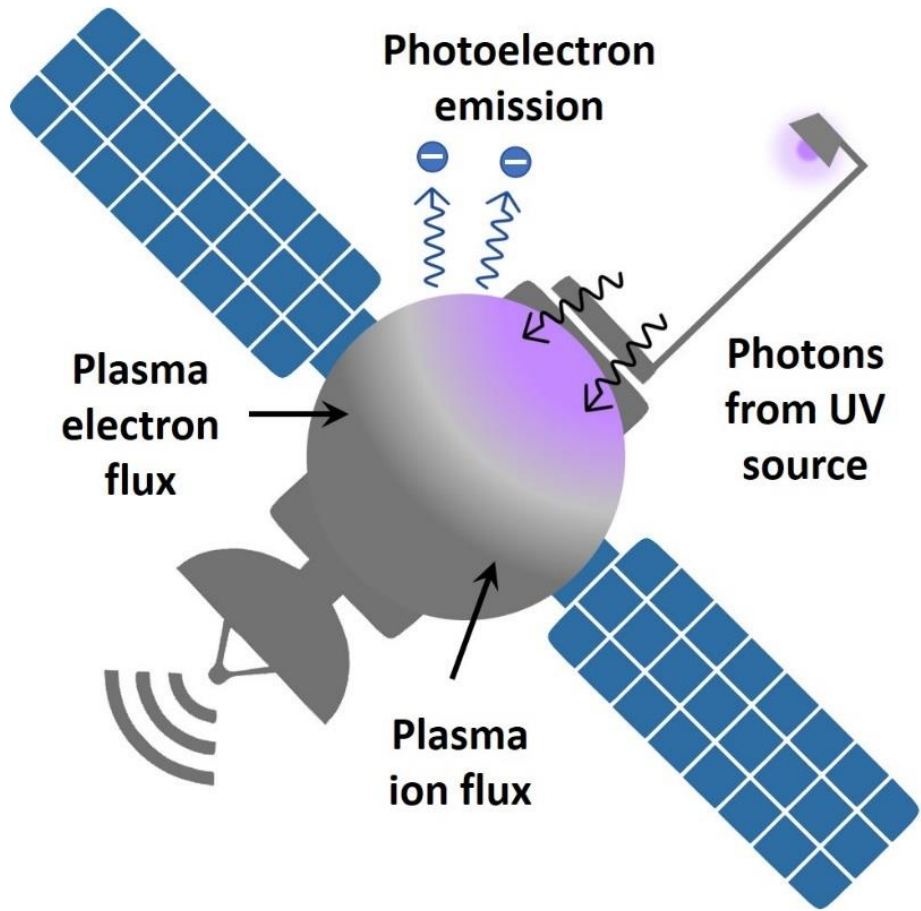

**Fig.4:** 1st mitigation scheme: A schematic illustrating the UV source assembly over satellite operating in the dark/ shadow/ eclipse region.

**Fig.5**

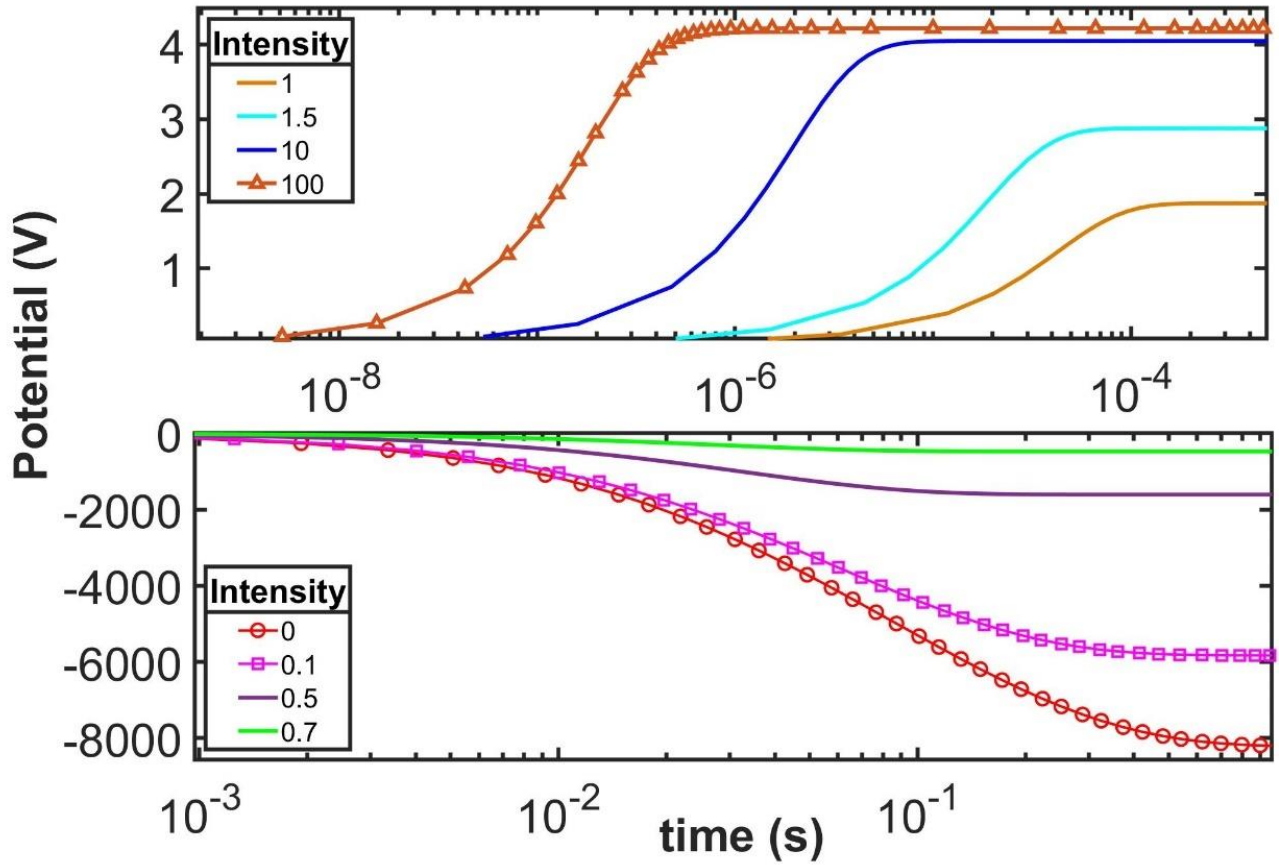

**Fig.5:** Temporal evolution of the surface potential of the satellite with an UV setup operating in the dark/ shaded region of GEO. The colored lines refer to the different illumination intensities.

**Fig.6**

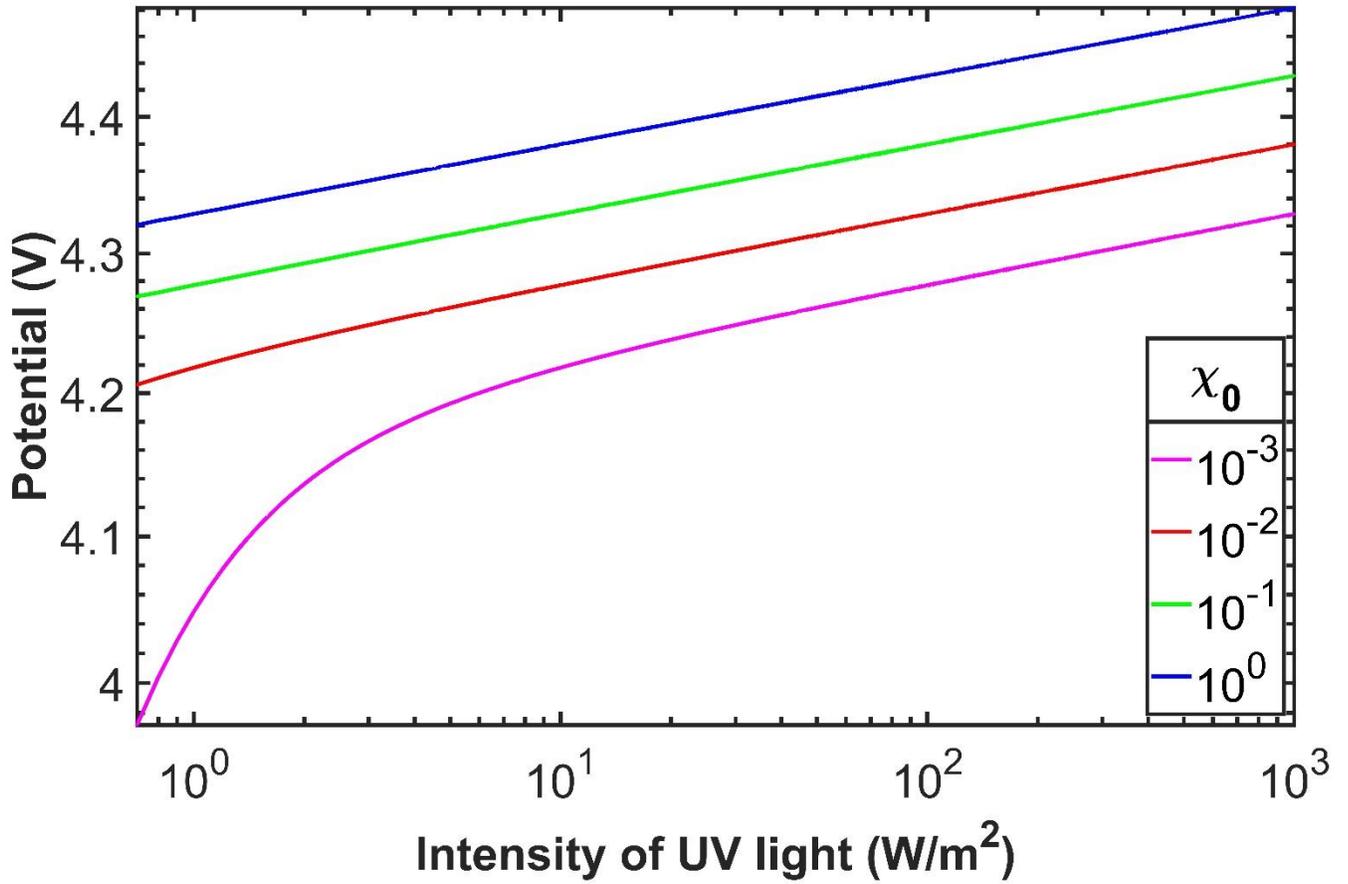

**Fig.6:** The dependence of steady state surface potential as a function of UV light intensity. The colored lines refer to the different values of the photoelectric yield ($\chi_0$).

**Fig.7**

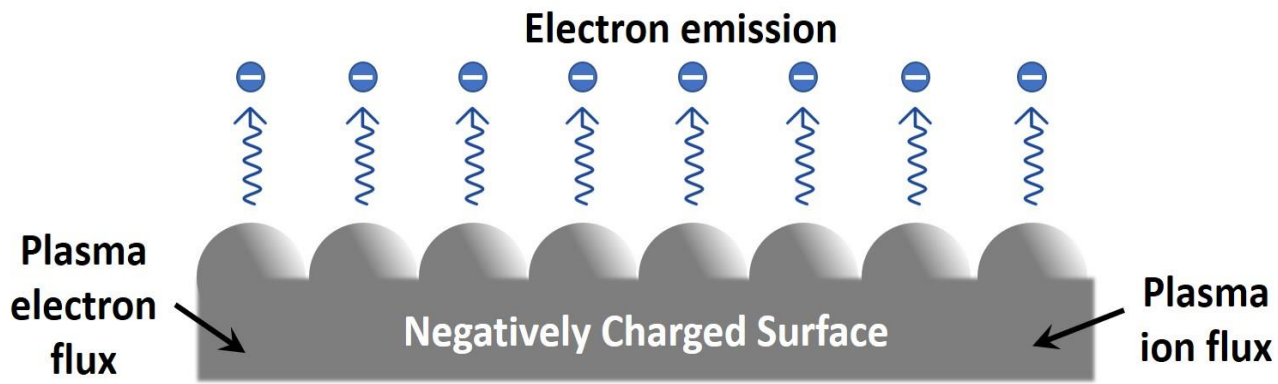

**Fig.7:** 2nd mitigation scheme: A schematic illustration of the nanofabricated surface of the satellite operating in the dark/ shadow/ eclipse region.

**Fig.8**

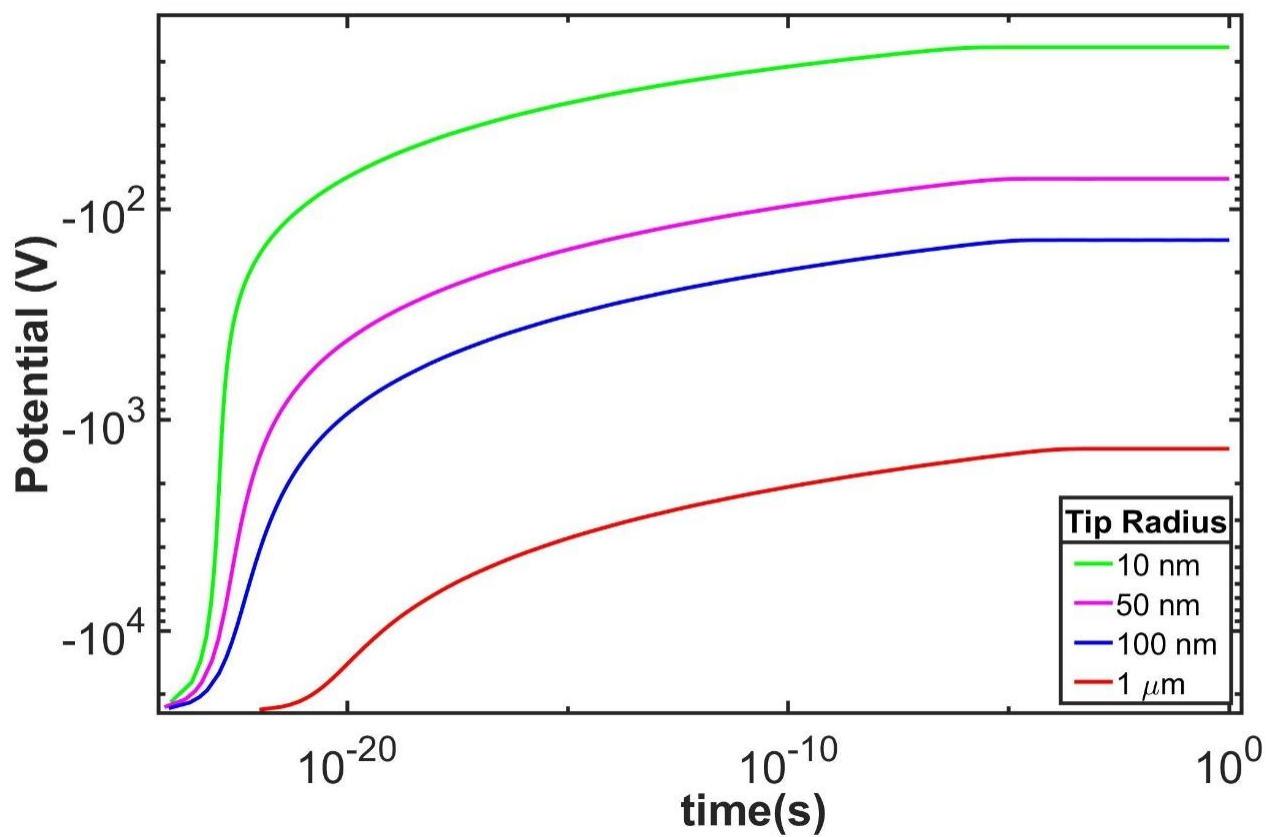

**Fig.8:** Temporal evolution of the electric potential on the satellite surface fabricated with the spherical tips, operating in the dark/ shaded region of the geomagnetic substorm. The colored lines refer to surfaces with different tip sizes.

**Fig.9**

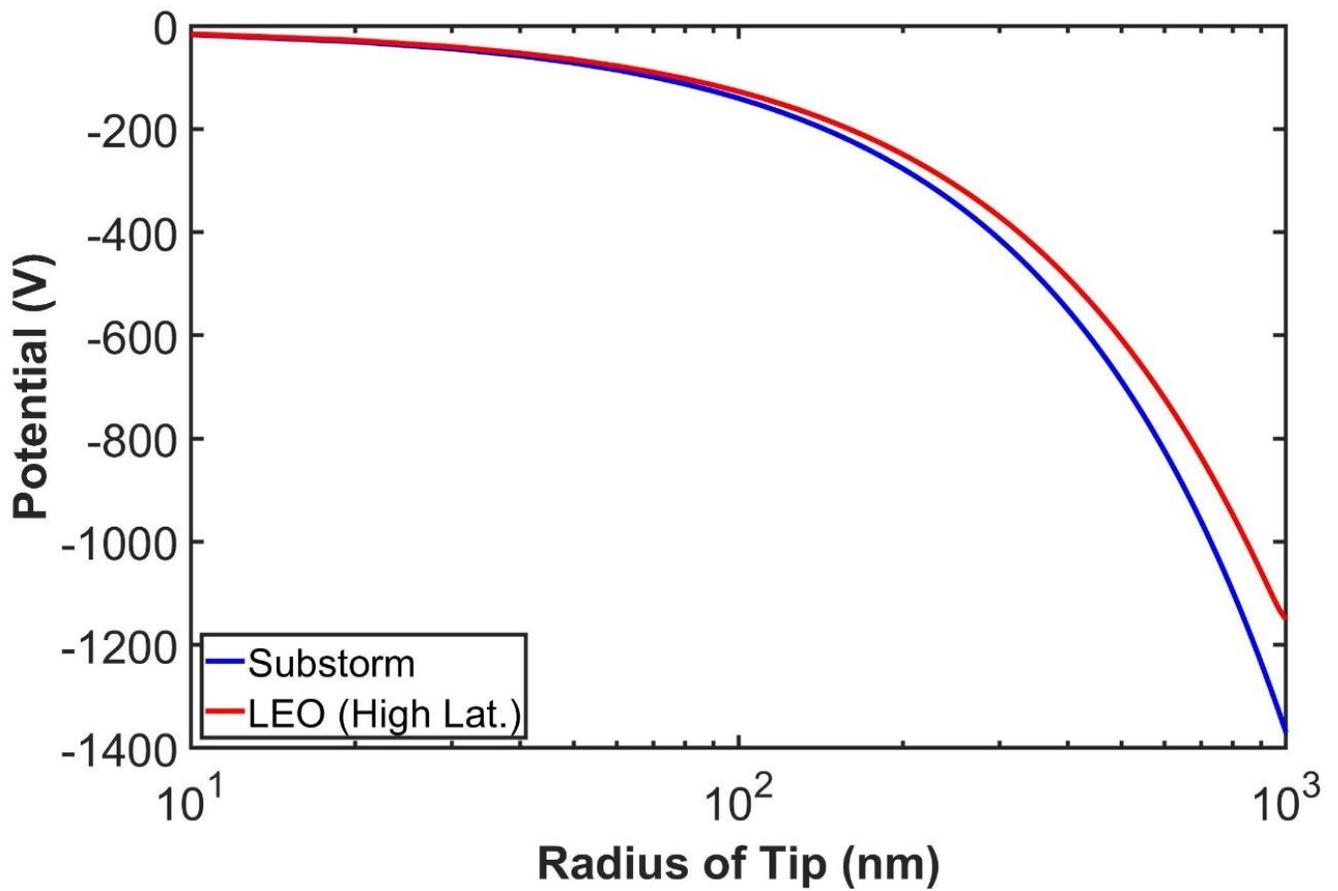

**Fig.9:** The dependence of steady state surface potential of the satellite as a function of the tip size of the fabricated surface operating in the dark/ shaded region. The colored lines refer to the different plasma conditions, viz., geomagnetic substorm and high latitude LEO regions.

---------------------------------------